\documentclass[11pt]{article}
\usepackage{epsfig}
\def \seff {s^2_{eff}}

\newcommand{\be}{\begin{equation}}
\newcommand{\ee}{\end{equation}}
\newcommand{\bea}{\begin{eqnarray}}
\newcommand{\eea}{\end{eqnarray}}
\newcommand{\smallz}{{\scriptscriptstyle Z}} 
\newcommand{\smallw}{{\scriptscriptstyle W}} %
\newcommand{\smallh}{{\scriptscriptstyle H}} %
\newcommand{\mz}{M_\smallz}
\newcommand{\mw}{M_\smallw}
\newcommand{\mh}{M_\smallh}
\newcommand{\mt}{M_t}
\newcommand{\gl}{\Gamma_\ell}
\newcommand{\equ}[1]{Eq.~(\ref{#1})}
\newcommand{\efe}[1]{Ref.\cite{#1}}
\newcommand{\lsim}{\;\rlap{\lower 3.5 pt \hbox{$\mathchar \sim$}} \raise 1pt
 \hbox {$<$}\;}
\newcommand{\gsim}{\;\rlap{\lower 3.5 pt \hbox{$\mathchar \sim$}} \raise 1pt
 \hbox {$>$}\;}

\begin{document}
\begin{titlepage}
\begin{flushright}
        \small
         DFPD-01/TH/05\\
         February 2001
\end{flushright}

\begin{center} 
\vspace{1cm}
\renewcommand{\thefootnote}{\fnsymbol{footnote}}
{\large\bf Where is the Higgs?\footnote{Talk presented at 
``50 Years of Electroweak Physics: A symposium in honor
of Professor Alberto Sirlin's 70th Birthday'', New York University,
New York, October 27-28, 2000}}

\vspace{0.5cm}
{$$\bf   G.~Degrassi$$ }
\setcounter{footnote}{0}
\vspace{.3cm}

{\it   Dipartimento di Fisica, Universit{\`a}
                  di Padova, Sezione INFN di Padova,\\ 
                  Via F.~Marzolo 8 , I-35131 Padua, Italy\\
\vspace{1.5cm} }

{\large\bf Abstract}
\end{center}
I discuss the theoretical uncertainties in the indirect Higgs mass 
determination.
I present the probability density function for the Higgs mass obtained
combining together the information from precision measurements with
the results from the direct search experiments carried out at LEP.
The probability that the Higgs weights less than 116 GeV comes out to be 
 around $35\%$  while the $95\%$ upper limit is located around 210-230 GeV.
\vspace{.2cm} 

\end{titlepage} 
\section{Introduction}
The last months of the year 2000 have seen a great excitement in the 
physics community because of a possible evidence at LEP of a Higgs
boson with a mass  $\mh \approx 115$ GeV. Unfortunately the shutdown of LEP
will not allow additional information on the Higgs to be collected in the 
near future.  Thus, it seems a good moment to try to review what we
(do not) know about the Higgs.  

On one side we have the impressive  amount 
of data collected at LEP, SLC, and the Tevatron that allow to probe the 
quantum structure of the Standard Model (SM), thereby  providing indirect 
information about the Higgs mass. On the other we have the result of the
direct searches performed at LEP with the excess of events reported in the 
combined analysis  of the four collaborations. 
The virtual Higgs effects are usually analyzed through a $\chi^2$  fit to 
the various precision observables that allows a $95\%$ Confidence Level (C.L.)
upper bound to be derived. The outcome of the searches used to be reported as 
a combined $95 \%$ C.L. lower bound, while today is more appropriate described
by the likelihood of the experiments. Given these two pieces of information
it is often said that one of the greatest achievement of LEP has been to
have pin down the Higgs mass between 113  (from the $95 \%$ C.L. lower bound
of the direct searches) and 170 GeV (from the $95 \%$ C.L. upper bound of the 
global fit to electroweak data). What I would like to discuss in this talk
is how confident we are in the fact that $113 \leq \mh \leq 170$ GeV addressing
two aspects of this problem: i) how solid is the $170$ GeV $95 \%$ C.L. upper 
bound? ii) What is the right way to combine the information from the
precision measurements with that from the searches to answer the simple 
question:``what is the probability that the Higgs mass is between, for example,
113 and 170 GeV?''

\section{Theoretical uncertainties in the indirect Higgs mass  determination}
\label{sect1}
As well known,  the vector boson self-energies are sensitive to
virtual Higgs effects and therefore the value of the Higgs mass
affects  the precision measurements like, for example, the ones made 
at peak of the $Z$. However, the Higgs behavior in the relevant electroweak
corrections is very  mild,  just logarithmic. 
To appreciate how much this logarithmic behavior makes hard to constrain
the Higgs let me consider a simple example. Let us take the effective
electroweak mixing angle, $\sin^2 \!\theta^{lept}_{eff} \equiv \seff$,
that is the most important quantity in the determination  of $\mh$, and 
write it as 
\be
\seff \sim (c_1 + \delta c_1) + (c_2 + \delta c_2) \log \, y; \quad
 y \equiv (\mh/100\, \mbox{GeV}) .
\label{e1}
\ee
In \equ{e1} I identify  the l.h.s.~with the experimental result
$\seff = (\seff)_0 \pm \sigma(\seff)$ while  in the r.h.s.~$\delta c_i$ 
represent the 
theoretical uncertainty in the corresponding coefficients connected to the fact
that we have computed $c_i$ in perturbation theory through  certain order
in the perturbative series and therefore we do not know their exact values 
because of higher order contributions.
From \equ{e1}\ one obtains 
\be
y = y_0\, \exp  \left[ -  \frac{\Delta_{th}}{c_2}
               \pm \frac{\sigma}{c_2} \right]; \quad
\Delta_{th} =  \delta c_1 + \delta c_2 \log\, y
\label{e2}
\ee
where $y_0$ is the value corresponding to $\delta c_1= \delta c_2 = \sigma=0$.
To see the effect of $\Delta_{th}$ in extracting 
$\mh$ I put $\sigma =0$ and take  
\be
\quad c_2 \sim \frac{\alpha}{2  \pi(c^2 - s^2)}
\left(\frac56 - \frac34 c^2 \right) \sim 5.5 \times 10^{-4};
\quad \Delta_{th} \sim  \pm 1.4 \times 10^{-4}
\label{e3}
\ee
where $s^2  \sim 0.23, \:
c^2 = 1 - s^2$. In \equ{e3} I estimate
$c_2$ through the Higgs leading behavior of the 
correction $\Delta \hat{r}$ relevant for $\seff$ \cite{si89}
while for $\Delta_{th}$, 
I take the value estimated in the 1995 CERN report
on `Precision calculation for the Z resonance' \cite{CERN} that was supposed 
to represent the uncertainty due to next-to-leading two-loop electroweak 
effects.  Inserting the values of \equ{e3} into
\equ{e2} yields $y \sim 1.29 \,y_0$. We see that a theoretical uncertainty
coming from two-loop unknown contributions (that are supposed to be not
even the dominant part) makes an error in the indirect determination of
the Higgs mass  of 29 \%! 

Eq.(\ref{e2}) tells us an important thing, 
namely that the error on $y$ depends on its central value $(y_0)$. 
This implies that not always shrinking the uncertainty in $\ln y$ reduces
the uncertainty on $y$. This is true only if the central value does not 
change, but this is not always the case (as we will see later). 
This consideration can be put in a more formal way noticing that if the 
logarithm of a quantity, $A\equiv \ln y$, is normally distributed, 
then the quantity
itself is  distributed according to a lognormal (see, e.g., 
\cite{JK} for the properties of this  distribution) whose 
standard deviation,  given by
$$
\sigma(y) = \left(
\exp{\left[2\,\mbox{E}[A]+2\,\sigma^2(A)\right]}-
\exp{\left[2\,\mbox{E}[A]+\sigma^2(A)\right]}\right)^{\frac{1}{2}},
$$
is a combination of the expected value and standard deviation of  its 
logarithm and therefore compensating effects can happen. 

The above example clearly tells us that to extract accurate indirect 
information on the Higgs one needs both very precise experiments and
a very good control of the theory side. This brings in the issue of
what  error we can associate to our theoretical predictions.
They are affected by  uncertainties  coming from two 
different sources: one that is called parametric and it is connected to the
error in the experimental inputs used in our predictions. The second one 
is called intrinsic and it is related to the fact that our knowledge of the
perturbative series is always limited, usually to the first few terms. 
Concerning 
parametric uncertainties, $\alpha(0), \, G_\mu$ and $\mz$ are very well
measured, while the top mass, $\mt$, and the strong coupling constant,
$\alpha_s$, are not so precisely known.
However, the scale of the weak interactions
is given by the mass of the intermediate vector bosons, so what actually 
matters in our predictions is not $\alpha(0)$ but $\alpha(\mz)$. The latter
contains the hadronic contribution to the photon vacuum polarization,
$(\Delta \alpha)_h$,
that cannot be evaluated in perturbation theory. Fortunately, one can use a 
dispersion relation to relate it  to the experimental
data on the cross section for $e^+ e^-$ annihilation into hadrons. 
In the recent years this subject has received a lot of attention with 
the appearance of several new
evaluations that have followed two main streams. On one path there are the 
most  phenomenological (ph.) analyses that rely on the use of  all the 
available experimental data on the
hadron production in $e^+ \,e^-$ annihilation and on perturbative QCD (pQCD)
for the high energy tail ($E \geq 40$ GeV) of the dispersion 
integral \cite{Jeg,Phen}. On the other the so called ``theory driven'' 
(t.d.) analyses that advocate the
use of pQCD down to energy scale of the order of tau mass, supplemented by
the use  of the experimental data in  regions,
like, for example, the threshold for the charmed mesons, where pQCD is not 
applicable \cite{jeger,DH,Mar}. I am not going to discuss the differences
in the various analyses (see Fred Jegerlehner's talk \cite{Fred}) but
I would like to point out few facts: i) all results are compatible with each 
other. The choice of one value instead of another is just a matter
of taste (or friendship). ii) The t.d.\ results have a smaller 
error with respect to the ph.\ ones but a lower central value.
iii) I can take two perfectly compatible numbers, one from the ph. analyses
like $(\Delta\alpha)_h = 0.02804\pm0.00065$ \cite{Jeg} and one from the t.d.\
ones like  $(\Delta\alpha)_h = 0.02761\pm0.00022$ \cite{Mar} and get a 
difference in the $95\%$ C.L.\ upper bound for the Higgs mass 
${\cal O}(50\,\mbox{GeV})$. Just to mention, it is the t.d.\ value,
that has a smaller error, that gives the higher upper bound.

I would like to emphasize that this uncertainty
has nothing to do with the blue band in the famous $\Delta \chi^2$ vs.~$\mh$
LEPEWWG plot.  There, the blue band represents, for a chosen value of
$(\Delta\alpha)_h$, the intrinsic uncertainty. 
There is no way to rigorously define the intrinsic uncertainty. The best
that can be done is to try to estimate it  by comparing
the output of the two codes TOPAZ0 \cite{TOP} and ZFITTER \cite{ZFit}, 
that now include up to two-loop next-to-leading terms,
when they are run enforcing the several built in options for resumming known 
effects. These options are supposed to mimic the size of unknown
higher order terms and the numerical spread in the outputs of the two codes
can be taken just as an order of magnitude of the unknown higher order
contributions. 
\section{Higgs mass inference from precision measurements}
I am  going to discuss now what can we learn about the Higgs from precision
measurements. In the spirit of the question raised in the Introduction,
I am not going to perform the standard $\chi^2$ analysis,
although clearly what I present  will be related to it, but following a
Bayesian approach I am going to construct 
$ f(m_H\,|\,\mbox{\it ind.})$, the p.d.f. of the Higgs
mass conditioned by this indirect information under the assumption of
the validity of the S.M. I would do it employing the three 
observables, $\seff$, $\mw$ and $\gl$. These quantities are the most 
sensitive to the Higgs mass and also very accurate measured.  
The most convenient way to approach the problem is to make use 
of the simple parameterization proposed in \efe{DGPS} and updated in
\efe{DG99} where $\seff,\, \mw$ and $\gl$ are written as functions of
$\mh$, $\mt$, $\alpha_s$ and  the hadronic contribution 
to the running of the electromagnetic coupling:
\begin{eqnarray}
s^2_{eff} &=& (s^2_{eff})_\circ + c_1A_1+c_2A_2-c_3A_3+c_4A_4\,, 
\label{eq:DG3}\\
M_W       &=& M_W^\circ-d_1A_1-d_5A_1^2-d_2A_2+d_3A_3-d_4A_4\, ,
\label{eq:DG4}\\
  \Gamma_\ell & = & \Gamma_\ell^\circ - g_1\, A_1 -g_5\, A_1^2- g_2\, A_2 +
 g_3\, A_3 - g_4\, A_4\, .  
\label{eq:DG5}    
\end{eqnarray} 
In the above equations 
$A_1\equiv \ln(M_H/100\,\mbox{GeV})$,
$A_2\equiv\left[(\Delta\alpha)_h/0.0280-1\right]$, 
$A_3\equiv\left[(\mt /175\,\mbox{GeV})^2-1\right]$
and 
$A_4\equiv\left[(\alpha_s(M_Z)/0.118-1\right]$;
$(s^2_{eff})_\circ$, $M_W^\circ$ and $\gl^\circ$ are (to excellent 
approximation) the theoretical results obtained at the reference point
$(\Delta\alpha)_h=0.0280$, $\mt=175$\,GeV, and $\alpha_s(M_Z)=0.118$ while the
values of the coefficients $c_i$, $d_i$ and $g_i$ are reported in 
Tables 3-5 of \efe{DG99} for three different renormalization scheme.
Formulae (\ref{eq:DG3}--\ref{eq:DG5}) 
are very  accurate for $75 \lsim\mh\lsim 350$ GeV with the
other parameters in the ranges
$170\lsim\mt\lsim 181$ GeV, $0.0273\lsim (\Delta\alpha)_h\lsim
0.0287$, $0.113\lsim \alpha_s(M_Z) \lsim 0.123$. In this case they reproduce 
the exact results of the calculations of 
Refs.(\cite{DGV,DGS,DG99}) with maximal errors of 
$\delta s^2_{eff}\sim 1\times 10^{-5}$,
$\delta\mw\lsim 1$ MeV and  $\delta\Gamma_\ell\lsim 3$ KeV, which are all very 
much below the experimental accuracy. Outside the above range, the
deviations increase but remain very small for larger Higgs mass, 
reaching about $3\times 10^{-5}$, 3 MeV, and 4 KeV
at $M_H=600$ GeV for $\seff$, $\mw$, $\Gamma_\ell$, respectively.

Formula (\ref{eq:DG3}) can be seen as providing an indirect measurement
of $A_1= [ \seff - (\seff)_\circ - c_2 A_2 + c_3 A_3 -c_4 A_4]/c_1$,
while (\ref{eq:DG4}) and (\ref{eq:DG5}) of the quantities 
$Y = \mw^\circ - \mw - d_2 A_2 + d_3 A_3 -d_4 A_4$ and
$Z= \gl^\circ - \gl- g_2 A_2 + g_3 A_3 -g_4 A_4 $, 
respectively, all three variables being described by Gaussian p.d.f.'s.
The $A_1$, $Y$ and $Z$  determination are clearly correlated, 
therefore one has to built a covariance matrix. This can be easily done
because formulae (\ref{eq:DG3}--\ref{eq:DG5}) are linear in the common
terms $\underline{X}\equiv \{\mt,\,\alpha_s(M_Z),\, (\Delta\alpha)_h\}$. 
The likelihood of these indirect measurements 
$\underline{\Theta} \equiv \{A_1,\, Y, \, Z \}$
is then a three dimensional correlated normal with covariance matrix
\be V_{ij} =\sum_l \frac{\partial \Theta_i}{\partial X_l}\cdot 
\frac{\partial \Theta_j}{\partial X_l} \cdot\sigma^2(X_l)
\label{eq:corrmat}
\ee
or 
\be
f(\underline{\theta} \,|\, \ln(m_H/100)) \propto e^{- \chi^2/2}
\label{eq: indlik}
\ee
where $\chi^2 = \underline{\Delta}^T {\bf V}^{-1} \underline{\Delta}$ with 
$\underline{\Delta}^T = \{ a_1 - \ln(m_H/100),\,
y - d_1 \ln(m_H/100) -d_5 \ln^2(m_H/100),\,
z - g_1 \ln(m_H/100) -g_5 \ln^2(m_H/100)\}$.
Using Bayes' theorem  the likelihood (\ref{eq: indlik}) can be turned
into a p.d.f. through  the choice 
of a prior. The natural choice is a uniform prior in $\ln(m_H)$, 
since it is well understood that radiative corrections measure
this quantity (for this reason the  likelihood (\ref{eq: indlik})
has been expressed in terms of $\ln(m_H)$). 
The uniform prior in $\ln(m_H)$ implies 
that $f(\ln(m_H/100) \,|\,\mbox{\it ind.})$ 
is just the normalized likelihood (\ref{eq: indlik}).
Thus, using standard probability calculus I can express the
result as a p.d.f. of $M_H$:
\be 
f(m_H\,|\,\mbox{\it ind.}) = \frac{ m_H^{-1}\, e^{- (\chi^2/2)}}
         {\int_{0}^\infty m_H^{-1}\, e^{- (\chi^2/2)}\, \mbox{d}m_H}.
\ee
\section{Including the constraint from the direct search}
Given $f(m_H\,|\,\mbox{\it ind.})$ 
a natural question to ask is  how this p.d.f. should
be modified in order to take into account the knowledge that the Higgs
boson has not been observed (that there is some indication for the Higgs)
at LEP\footnote{A related 
discussion can be found in \efe{Er}.}. To answer this 
question let me discuss first an ideal case.
I consider a search for Higgs production in association with a particle of 
negligible width in an experimental  situation of ``infinite'' luminosity,
perfect efficiency and   no background whose outcome was no candidate. 
In this situation we are sure that  all mass values below a  sharp kinematical 
limit $M_K$ are excluded. 
This implies that: a) the p.d.f. for $M_H$  must vanish below $M_K$;
b) above $M_K$ the relative probabilities cannot change, because 
there is no sensitivity in this region, and then the experimental
results cannot give information over there. 
For example, if $M_K$
is $110$ GeV, then $f(200\, \mbox{GeV})/f(120\, \mbox{GeV})$ 
must remain constant before and after 
the new piece of information is included. 
In this ideal case we have then
\begin{equation}
f(m_H\,|\,\mbox{\it dir.} \,\&\,\mbox{\it ind.}) = 
\left\{\begin{array}{ll} 
0   & m_H < M_K  \\
 \frac{f(m_H\,|\,\mbox{\it ind.})}
    {\int_{M_K}^\infty f(m_H\,|\,\mbox{\it ind.})\, \mbox{d}m_H} & m_H \ge M_K\,, 
\end{array}\right.
\label{eq:taglio_secco} 
\end{equation}   
where the integral at denominator is just a normalization coefficient. 

More formally, this result can be obtained making explicit use
of the Bayes' theorem. Applied to our problem, the theorem
can be expressed as follows (apart from a
 normalization constant):
\begin{equation}
f(m_H\,|\, \mbox{\it dir.} \,\&\,\mbox{\it ind.})
 \propto f(\mbox{\it dir.}\,|\,m_H)\cdot  f(m_H\,|\,\mbox{\it ind.})\,,
\label{eq:Bayes}
\end{equation} 
where $f(dir\,|\,m_H)$ is the so called  likelihood. 
In the idealized example  we are considering now, 
$f(dir\,|\,m_H)$ can be expressed in terms of
the probability of  observing zero candidates in an experiment sensitive 
up to a $M_K$ mass for a given value $m_H$, or
\begin{equation}
f(\mbox{\it dir.} \,|\, m_H) = f(\mbox{``zero cand.''}\,|\,m_H) = 
\left\{\begin{array}{ll} 
0   & m_H < M_K \\
      1 & m_H \ge M_K \, . 
\end{array}\right.
\label{eq:lik_taglio_secco} 
\end{equation}   
In fact, we would expect an ``infinite'' number of events 
if $M_H$ were below the kinematical limit.
Therefore the probability of observing nothing should be zero. 
Instead, for $M_H$ above $M_K$,
the condition of vanishing production cross section and no background
can only yield no candidates. 

Consider now a real life situation.  In this case 
the transition between Higgs mass values which are impossible to 
those which are possible is not so sharp. 
In fact because of physical reasons (such as threshold effects
and background) and experimental reasons (such as luminosity
and efficiency) we cannot be  really 
sure about excluding  values close to the kinematical limit, 
nevertheless the ones very far from $M_K$  are ruled out.
Furthermore, the kinematical limit is in general not sharp; at LEP,
for example, the large total width of the $Z^\circ$ plays an important
role.  
Thus, in a  real life  situation we expect  the ideal step function 
likelihood of  \equ{eq:lik_taglio_secco} to be replaced by a smooth curve 
which goes to zero for low masses. Concerning, instead, the region of no 
experimental sensitivity, $M_H \gsim M_{K_{eff}}$, the likelihood is 
expected
to go to a value  independent on the Higgs mass that however is different
from that of the ideal case, i.e.\ 1, because of the presence of the 
background.

In order to  combine the various pieces of information easily it is  
convenient to replace the likelihood by a function 
that goes to 1 where the experimental sensitivity is 
lost \cite{dd,pia_giulio}. Because constant factors do not play any role in 
the 
Bayes' theorem this can be achieved by dividing the likelihood by its value
calculated for very large Higgs mass values where no signal 
is expected, i.e. the case of pure background. This likelihood ratio, 
${\cal R}$, can be seen as the counterpart, in the case of a real 
experiment, of the step function of \equ{eq:lik_taglio_secco}. 
Therefore, the Higgs mass p.d.f. that takes into account both 
direct search and precision measurement results can be written as 
\be 
f(m_H\,|\,\mbox{\it dir.} \,\&\,\mbox{\it ind.}) =
\frac{ {\cal R}(m_H)\, f(m_H\,|\,\mbox{\it ind.})}
    {\int_{0}^\infty {\cal R}(m_H)\, f(m_H\,|\,\mbox{\it ind.})\,\mbox{d}m_H}~.
\label{fine}
\ee
In \equ{fine} ${\cal R}$, namely the information from the direct searches,
acts as a shape distortion function of $f(m_H\,|\,\mbox{\it ind.})$. 
As long as ${\cal R}(m_H)$ 
is 1, the shape (and therefore the relative probabilities in that region)
remains unchanged, while ${\cal R}(m_H)\rightarrow 0$ indicates regions 
where the p.d.f. should vanish. 

One should notice that 
${\cal R}(m_H)$ can also assume values larger than 1 for Higgs mass values
below the kinematical limit. This situation corresponds to a number of 
observed candidate events larger than the expected background. In this case 
the role played by
${\cal R}(m_H)$ is to stretch $f(m_H\,|\,\mbox{\it ind.})$ 
below the effective kinematical limit and this might even
prompt a claim for a discovery if ${\cal R}$ becomes sufficiently large 
for the probability of $M_H$ in that region to get 
very close to 1.
\section{Results}
The experimental inputs  I use to construct 
$f(m_H\,|\,\mbox{\it ind.})$
are \cite{LEPEWWG}: $\seff=0.23146\pm 0.00017$, $\mw=80.419\pm 0.038$ GeV, 
$\Gamma_\ell=83.99\pm 0.10$ MeV, $\mt=174.3\pm 5.9 $ GeV,
$\alpha_s(M_Z)=0.119\pm0.003$. Concerning  $(\Delta\alpha)_h$, as I said
we have many possible choices. I am going to present my results using
two different values of $(\Delta\alpha)_h$, one as representative of
the ph. analyses  and the other for the t.d.\ evaluations. For the ph.\ case
I take the standard value 
$(\Delta\alpha)_h^{EJ} = 0.02804\pm0.00065$~\cite{Jeg}
while for the t.d.\ analyses I choose the one that has the smallest
uncertainty, i.e. $(\Delta\alpha)_h^{DH}=0.02770\pm0.00016\,$~\cite{DH}. 
The intrinsic uncertainty is taken into account by averaging different
inferences (see \efe{dd}). 

The values of the ${\cal R}$ function that enters in \equ{fine} has been
provided by the LEP Higgs Working Group \cite{Igo}: they take into account
all Higgs searches  by the  four LEP collaborations.
\begin{table}[t]
\begin{center}
\begin{tabular}{|l|cc|cc|}\hline
& & & &   \\
  & \multicolumn{2}{|c|}{$(\Delta\alpha)_h= 0.02804(65)$} 
  & \multicolumn{2}{|c|}{$(\Delta\alpha)_h= 0.02770(16)$} \\
& & & &   \\ \hline
& & & &   \\
& $ (ind.)$ 
& $\left(\begin{array}{c} ind. \\ + \\ dir. \end{array}\right)$ 
&$(ind.)$ 
& $\left(\begin{array}{c} ind. \\ + \\ dir. \end{array}\right)$\\
& & & &   \\
$\mbox{E}[M_H]$/GeV  &  90   & {\bf 140}  &  105 &{\bf 135}\\
$\sigma(M_H)$/GeV        &  55   & {\bf  45}  &   50 &{\bf  35}\\
& & & &   \\
$P(M_H\le 110\,\mbox{GeV})$          &73\,\% & $\sim 0$ & 61\,\% & $\sim 0$ \\
$P(M_H\le 116\,\mbox{GeV})$          &76\,\% & 37\,\%  & 66\,\% & 38\,\% \\
$P(M_H\le 130\,\mbox{GeV})$          &82\,\% & 66\,\%  & 75\,\% & 68\,\%  \\
$P(M_H\le 200\,\mbox{GeV})$          &95\,\% & 91\,\%  & 95\,\% & 94\,\%  \\
& & & &   \\
$M^{95}_H$/GeV; $P(M_H\le M^{95}_H)\approx 0.95$ 
                                          &195 &{\bf 230}& 200& {\bf 210}\\
$M^{99}_H$/GeV; $P(M_H\le M^{99}_H)\approx 0.99$ 
                                          &290 &{\bf 330}& 275& {\bf 280}\\
& & & &   \\
\hline  
\end{tabular}
\end{center}
\caption{\sf Summary of the direct plus indirect information.}
\label{tab:results}
\end{table}

\begin{figure}[t]
\begin{center}
\epsfig{file=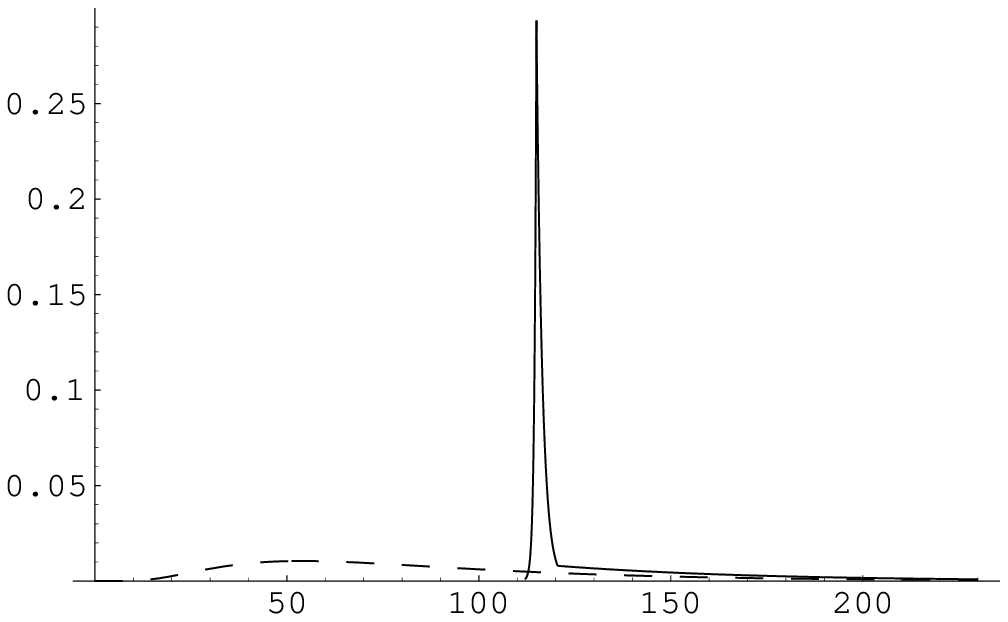,width=0.60\linewidth}\\[0.2cm]
\epsfig{file=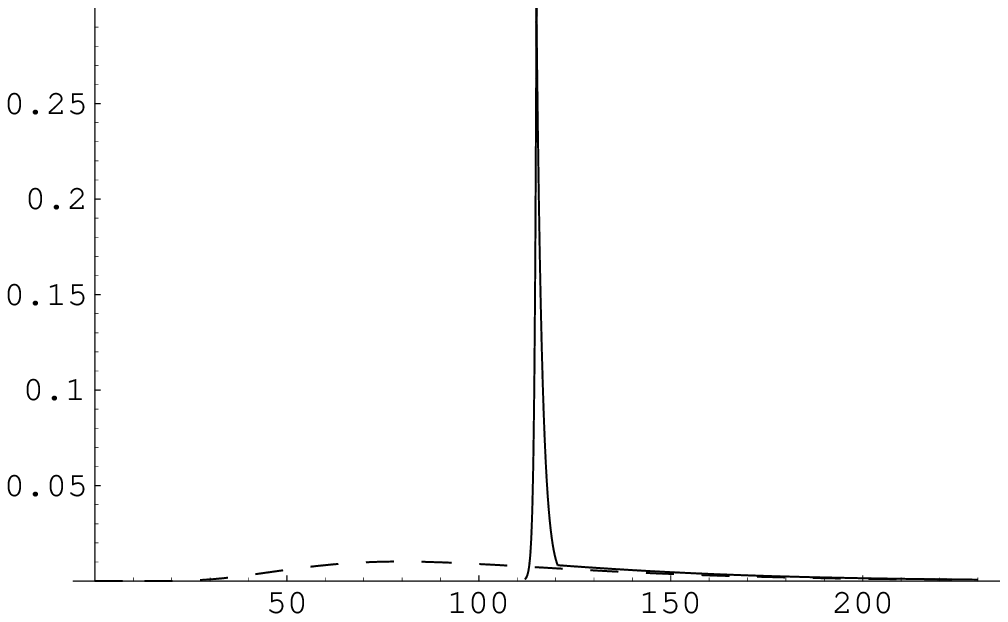,width=0.60\linewidth}
\end{center}
\caption{\sf Probability distribution functions using only indirect information
(dashed line) and employing also the experimental results from direct searches
(solid one): a) $(\Delta \alpha)_h = 0.02804(65)$; b) 
$(\Delta \alpha)_h = 0.02770(16)$.}

\vspace{-11.2cm}\hspace{+3.8cm}\mbox{\scriptsize{$f(M_H)$}}

\vspace{+0.5cm}\hspace{+7.8cm}\mbox{a)}

\vspace{+2.0cm}\hspace{+10.2cm}\mbox{\scriptsize{$M_H$}}

\vspace{+1.4cm}\hspace{+3.8cm}\mbox{\scriptsize{$f(M_H)$}}

\vspace{+0.5cm}\hspace{+7.8cm}\mbox{b)}

\vspace{+1.9cm}\hspace{+10.2cm}\mbox{\scriptsize{$M_H$}}

\vspace{2.4cm}
\label{fig:Jeg}
\end{figure}

Table \ref{tab:results} summarizes the result of my analysis.
The shape  of the
p.d.f. with and without the inclusion of the direct search information
is presented in Fig.~\ref{fig:Jeg}.
From this figure one can  notice that, in the case of  
$f(m_H\,|\, \mbox{\it ind.} )$, the use
of a higher central value for $(\Delta\alpha)_h$ 
(i.e.~$(\Delta\alpha)_h^{EJ}$)\ tends to concentrate more the
probability towards smaller values of $M_H$. As a consequence the 
analysis based on $(\Delta\alpha)_h^{EJ}$  gives
results for the expected value, E[$\mh$], 
the standard deviation, $\sigma(\mh)$,  of the  $M_H$ p.d.f. and 
95\,\% probability upper limit, $M_H^{95}$, very close to those obtained
using $(\Delta\alpha)_h^{DH}$ regardless the fact that the uncertainty
on $(\Delta\alpha)_h^{EJ}$ is 
approximately $4$ times larger than that of  $(\Delta\alpha)_h^{DH}$.
As expected, the inclusion  of the direct search information 
drifts the p.d.f.~towards higher
values of $M_H$ by cutting regions below 110 GeV. However, the most
striking thing in the figure is the spike around $\mh \approx 115$ GeV.
Indeed, due to the excess of events in the combined data set of the
LEP experiments, the ${\cal R}$
function  around 115 GeV takes values $\sim 30$ strongly enhancing the 
probability of that region.
\section{Conclusions}
The analysis that I have presented clearly shows that a heavy
Higgs scenario is highly disfavored, given a ${\cal O}(90\%)$
probability  that the Higgs weights less than 200 GeV.
However, it should be said that the distribution has a quite long tail.
Indeed we have that above 300 GeV there is still a residual 
${\cal O}(1\%)$ probability.
The  excess in the data recorded by the LEP collaborations is reflected in 
the analysis by the spike in the distribution around $\mh \approx 115$ GeV.
Maybe it can be of some interest to know that, according to my
analysis, the probability that the Higgs mass is below 116 GeV
is $37 \%$, if we use
$(\Delta\alpha)_h^{EJ}$, and $38 \%$ in the case of $(\Delta\alpha)_h^{DH}$.

Let me say clearly that all the results I have presented are derived
under the assumptions of the validity of the SM and rely on the experimental
inputs I have used. In particular I have used for $\seff$ the combined
LEP+SLD value, although it is well known that the two most precise 
determinations of it are not in very good agreement. Just to see how much my 
conclusions  rely on the combined $\seff$ value I have
redone the analysis discarding the SLD result, i.e.\ using the LEP
value for the effective sine, $\seff = 0.23184 \pm 0.00023$. The
results of this  extreme exercise are not very different: in the worst
case, i.e. employing $(\Delta\alpha)_h^{DH}$, 
$P(M_H\le 200\,\mbox{GeV})= 75\,\%$ while $M^{95}_H= 320$ GeV.
Finally there is still a $23 \%$ probability that the Higgs mass is below
116 GeV.

\vspace{1cm}
I am grateful to  G. D'Agostini, S.~Fanchiotti, F.~Feruglio,
 P.~Gambino, M.~Passera, A.~Sirlin
and A.~Vicini for the fruitful collaboration on  different
projects variously connected with the subject discussed here.
I would like also to thank M.~Porrati and the other organizers for the 
excellent organization and the very pleasant atmosphere of the workshop.
This work was supported in part by the  European Union under the contract
HPRN-CT-2000-00149.

\end{document}